\documentclass[preprint,12pt]{elsarticle}

\usepackage{amsmath,amssymb,amsfonts}
\usepackage{graphicx}
\usepackage{subcaption}
\usepackage{booktabs}
\usepackage{multirow}
\usepackage{tikz}
\usepackage{pgfplots}
\pgfplotsset{compat=1.17}
\usepackage{hyperref}
\usepackage{cleveref}
\usepackage{float} % For [H] placement
\usepackage{xcolor} % For coloring in TikZ
\usepackage{tabularx} % For flexible table width

\usetikzlibrary{shapes.geometric, arrows.meta, positioning, fit, backgrounds, calc}

\newcommand{\Conv}{\text{Conv}}
\newcommand{\Linear}{\text{Linear}}
\newcommand{\LayerNorm}{\text{LayerNorm}}
\newcommand{\GELU}{\text{GELU}}
\newcommand{\Softmax}{\text{Softmax}}
\newcommand{\GAP}{\text{GAP}}
\newcommand{\Encoder}{\text{Encoder}}
\newcommand{\Decoder}{\text{Decoder}}
\newcommand{\Patchify}{\text{Patchify}}

\newcommand{\MSE}{\text{MSE}}
\newcommand{\CE}{\text{CE}}

\usepackage[T1]{fontenc}
\usepackage[utf8]{inputenc}
\usepackage{textcomp} % For \textunderscore

\begin{document}

\begin{frontmatter}

\title{Entropy-Guided Self-Supervised Learning for Medical Image Classification}

\author[inst1]{Joao Florindo}
\ead{florindo@unicamp.br}
\author[inst1]{Viviane Moura}
\ead{v215939@dac.unicamp.br}

\address[inst1]{Institute of Mathematics, Statistics and Scientific Computing\\ 
	Department of Applied Mathematics, University of Campinas\\ Rua Sergio Buarque de Holanda, 651, Campinas, 13083-859, Sao Paulo, Brazil}

\begin{abstract}
Accurate and robust medical image classification is paramount for early disease diagnosis and treatment planning. However, challenges such as limited annotated data, high intra-class variability, and subtle inter-class differences often hinder the performance of deep learning models. This paper introduces a synergistic deep learning framework that leverages the strengths of self-supervised learning and transfer learning for enhanced medical image classification. Our approach employs two distinct ConvNeXt-Tiny models: one pre-trained on a large-scale natural image dataset (ImageNet) and another pre-trained using an entropy-guided Masked Autoencoder (MAE) on the target medical dataset. Both models are then fine-tuned on specific medical image classification tasks. A final ensemble strategy, based on averaging predicted probabilities, is utilized to combine the complementary insights from these two models. Rigorous experimental validation across four diverse medical imaging datasets (Breast Ultrasound Images (BUSI), International Skin Imaging Collaboration (ISIC) 2018, Kvasir, and COVID) demonstrates the superior performance and robustness of our ensemble approach. The MAE pre-training significantly improves feature learning on domain-specific data, while the ImageNet pre-training provides strong generalizable features. The ensemble consistently achieves state-of-the-art results, outperforming individual models and existing methods, highlighting the efficacy of combining diverse pre-training strategies for challenging medical image analysis.
\end{abstract}


\end{frontmatter}

\section{Introduction}
Medical image classification plays a pivotal role in modern healthcare, enabling automated or semi-automated diagnosis of various diseases from diverse imaging modalities such as X-rays, CT scans, MRIs, and ultrasound images \cite{yamashita2018convolutional, atasever2023comprehensive}. The ability to accurately and efficiently classify medical images can significantly assist clinicians in early detection, treatment planning, and monitoring disease progression, thereby improving patient outcomes \cite{jiang2023review, yellu2024medical}.

Deep learning, particularly Convolutional Neural Networks (CNNs), has revolutionized computer vision and subsequently medical image analysis \cite{lecun1989backpropagation, simonyan2014very, he2016deep}. CNNs excel at extracting hierarchical features, from low-level textures to high-level semantic concepts, due to their inherent inductive biases such as translation invariance and local receptive fields \cite{lowe2004distinctive}. However, medical imaging datasets often present unique challenges: they are typically smaller than natural image datasets, highly imbalanced, contain subtle yet critical features, and are susceptible to noise and artifacts \cite{salehi2023study, spolaor2024fine}. These characteristics can limit the generalization capabilities of CNNs trained from scratch or even those fine-tuned from ImageNet pre-trained weights, as ImageNet features may not fully capture domain-specific nuances.

In recent years, Vision Transformers (ViTs) \cite{dosovitskiy2020image} have emerged as powerful alternatives, demonstrating superior performance in various computer vision tasks by leveraging self-attention mechanisms to capture long-range dependencies. Hybrid CNN-Transformer architectures have also gained traction, aiming to combine the local feature extraction strengths of CNNs with the global contextual understanding of Transformers \cite{guo2022cmt, lin2023scale}. However, ViTs often require vast amounts of data for effective training, and their computational complexity can be a bottleneck, especially for high-resolution medical images \cite{liu2021swin, liu2022convnet}.

Self-supervised learning (SSL) has shown immense promise in mitigating the data scarcity problem in medical imaging by pre-training models on unlabeled data to learn rich, transferable representations \cite{azizi2021big, nielsen2023self}. Masked Autoencoders (MAEs) \cite{he2022masked} are a prominent SSL technique that involves masking a large portion of input patches and reconstructing the missing pixels. This forces the model's encoder to learn robust representations from partial observations, which can then be fine-tuned for downstream tasks. 

Inspired by ``diffusion forcing'' \cite{chen2024diffusion}, here we propose the ``entropy-guided'' masked autoencoder. This essentially consists in computing an entropy score for each patch, and then independently applying Gaussian noise over each patch with the Gaussian variance controlled by the entropy score. This turns the masking process into a continuous transform. The extreme case of maximum noise would be equivalent to the classical masking in MAE. The ``entropy-guided'' aspect of MAE pre-training, as explored in this work, further refines the masking strategy to focus on regions with higher information content or uncertainty, potentially leading to more discriminative feature learning for medical images.

Overall, this paper proposes a synergistic deep learning framework for medical image classification that integrates the benefits of ImageNet pre-training, entropy-guided MAE pre-training, and ensemble learning. Our approach utilizes two ConvNeXt-Tiny models \cite{liu2022convnet} as backbone architectures. One model is initialized with weights pre-trained on ImageNet, providing a strong baseline with generalizable features. The second model undergoes an entropy-guided MAE pre-training phase on the target medical dataset, enabling it to learn domain-specific, robust representations from unlabeled data. Both models are then fine-tuned on the labeled medical image classification task. Finally, an ensemble strategy, which averages the predicted probabilities from these two fine-tuned models, is employed to leverage their complementary strengths and enhance overall classification performance and robustness.

Our main contributions are summarized as follows:
\begin{itemize}
	\item We introduce an entropy-guided MAE pre-training strategy designed to learn domain-specific, robust feature representations from medical images by focusing on informative regions during masking.
    \item We propose a novel synergistic framework for medical image classification that combines ImageNet pre-training, entropy-guided MAE pre-training, and ensemble learning to achieve superior performance and robustness.
    %\item We detail the architecture and mathematical formulation of the ConvNeXt-Tiny backbone, its application in both ImageNet transfer learning and entropy-guided MAE self-supervised learning.
    \item We conduct extensive experiments on four diverse medical imaging datasets (BUSI, ISIC2018, Kvasir, COVID), demonstrating that our SSL approach consistently outperforms individual models and achieves state-of-the-art results across various metrics.
    \item We provide a comprehensive ablation study analyzing the impact of different pre-training strategies and hyperparameter settings on the overall performance.
\end{itemize}

The remainder of this paper is organized as follows: Section \ref{sec:related_works} reviews related works. Section \ref{sec:proposed_method} describes the proposed methodology and architectural details. Section \ref{sec:experimental_setup} outlines the experimental setup. Section \ref{sec:results} presents and analyzes the experimental results and ablation studies. Section \ref{sec:sota_comparison} provides a comparison with state-of-the-art methods. Finally, Section \ref{sec:conclusion} concludes the paper.

\section{Related Works}
\label{sec:related_works}
Medical image classification has witnessed significant advancements with the advent of deep learning, yet it continues to present unique challenges due to the inherent characteristics of medical data. This section reviews key developments in deep learning for medical image classification, focusing on CNN-based methods, Transformer-based methods, hybrid architectures, and self-supervised learning approaches, while integrating recent literature and the provided auxiliary documents.

\subsection{CNN-based Methods}
Convolutional Neural Networks (CNNs) have long been the cornerstone of medical image analysis due to their ability to automatically extract hierarchical features and their inductive biases like translation invariance and locality \cite{lecun1989backpropagation, simonyan2014very}. Early architectures like VGG \cite{simonyan2014very} and ResNet \cite{he2016deep} demonstrated the power of increasing depth and residual connections for improved performance. DenseNet \cite{huang2017densely} further enhanced feature reuse through dense connectivity, reducing parameter count and improving generalization. More recently, ConvNeXt \cite{liu2022convnet} re-evaluated modern CNN design principles, incorporating techniques from Vision Transformers such as Layer Normalization and depthwise separable convolutions, achieving strong performance while maintaining CNN efficiency.

In medical imaging, CNNs have been widely applied for tasks such as disease diagnosis and lesion localization. For instance, ResNet-based models have been used for breast cancer histological image classification \cite{jiang2019breast}. DAFNet \cite{cai2026dafnet} proposes a Dynamic Adaptive Fusion Network that enhances feature extraction and classification performance across diverse medical imaging tasks by integrating channel shuffling, feature transformation, adaptive normalization convolution (ANC), and dynamic attention mechanisms. This lightweight CNN architecture demonstrates superior classification accuracy and cross-domain adaptability. Similarly, Conv-SdMLPMixer \cite{ren2025conv} introduces a hybrid network combining CNNs and Multilayer Perceptrons (MLPs) to address issues of high noise and small lesion areas in medical datasets. It employs a multi-path inverted residual bottleneck CNN and a multi-scale multi-dimensional feature fusion MLP to capture both local details and global context, achieving outstanding performance on datasets like BUSI and COVID19-CT.

\subsection{Transformer-based Methods}
Originating from natural language processing \cite{vaswani2017attention}, Transformer models, particularly Vision Transformers (ViTs) \cite{dosovitskiy2020image}, have made significant inroads into computer vision. ViTs process images as sequences of patches, leveraging self-attention to model long-range dependencies. While powerful, ViTs typically require large-scale datasets for effective training and may struggle with local feature extraction due to the lack of inherent inductive biases present in CNNs \cite{liu2021swin}.

In medical imaging, Transformers are increasingly being explored. MedViT \cite{manzari2023medvit, manzari2025e} proposes a robust Vision Transformer for generalized medical image classification, integrating KANs (Kolmogorov-Arnold Networks) and Dilated Neighborhood Attention to enhance feature extraction and contextual understanding. Other works like MedMamba \cite{yue2024k} explore Vision Mamba architectures for medical image classification, combining the strengths of state space models with vision transformers. A distillation approach to Transformer-based medical image classification with limited data has also been proposed to address data scarcity \cite{sevinc2025distillation}.

\subsection{Hybrid Architectures}
Recognizing the complementary strengths of CNNs and Transformers, hybrid architectures have emerged to leverage both local and global feature extraction. CMT \cite{guo2022cmt} combines CNNs for local feature extraction with Transformers for long-range dependencies. SMT \cite{lin2023scale} replaces conventional convolution with a convolutional modulation module for multi-scale feature extraction and information fusion. CTransCNN \cite{wu2023ctranscnn} integrates multi-label multi-head attention with multi-branch residual modules for enhanced feature transfer.

HiFuse \cite{huo2024hifuse} proposes a hierarchical multi-scale feature fusion network that effectively fuses multi-scale global and local features without destroying their respective modeling capabilities. It employs a three-branch parallel structure with global and local feature blocks and an adaptive hierarchical feature fusion block (HFF block) to capture rich semantic and spatial information. Similarly, EFFResNet-ViT \cite{hussain2025effresnet} presents a fusion-based convolutional and Vision Transformer model for explainable medical image classification. A hybrid fully convolutional CNN-Transformer model for inherently interpretable medical image classification has also been explored \cite{djoumessi2025hybrid}. These models aim to achieve a better balance between local detail and global context, crucial for complex medical image analysis. The Central-Peripheral Vision Transformer (CPVT) \cite{lu2026biologically} is another biologically inspired hybrid CNN-Transformer architecture that mimics foveal and peripheral vision for medical image classification, achieving state-of-the-art performance on datasets like ISIC-2018 and Kvasir.

\subsection{Self-Supervised Learning and Other Advanced Architectures}% and Explainable AI}
Self-supervised learning (SSL) has become a vital paradigm for medical image analysis, particularly in scenarios with limited labeled data. Masked Autoencoders (MAEs) \cite{he2022masked} are a prominent SSL technique where models learn to reconstruct masked parts of an input, thereby acquiring rich representations. MAEs have been adapted for medical image classification to learn occlusion-invariant features \cite{kong2023understanding, xu2022self}. Medical Supervised Masked Autoencoder (MSMAE) \cite{mao2025medical} proposes a supervised attention-driven masking strategy (SAM) to precisely pinpoint main diseased tissues during pre-training, addressing the issue of subtle lesion regions being overlooked by high-ratio random masking.

%Explainable AI (XAI) is gaining increasing importance in medical imaging to build trust and provide insights into model decisions \cite{ukwuoma2025enhancing, ullah2025novel}. Methods like LIME and SHAP are used to enhance histopathological medical image classification for early cancer diagnosis \cite{ukwuoma2025enhancing}. A novel XAI framework integrating statistical, visual, and rule-based methods has been proposed for medical image classification \cite{ullah2025novel}. 

The Probabilistic Margin-Aware Focal Loss (PMAF Loss) \cite{sagar2025pmaf} is a unified objective function designed to enhance discriminability, robustness, and reliability in medical image classification by integrating focal modulation, uncertainty weighting, and margin-based feature regularization. Furthermore, advanced techniques like Diffusion Models are being explored. DiffMIC-v2 \cite{yang2025diffmic} proposes a diffusion-based network for medical image classification that eliminates unexpected noise and perturbations. It uses an improved dual-conditional guidance strategy and a novel heterologous diffusion process for efficient visual representation learning. Other works investigate adaptive dual-axis style-based recalibration networks for imbalanced medical image classification \cite{zhang2025adaptive} and multi-scale Transformer architectures \cite{hu2025multi}. A comparative study of statistical, radiomics, and deep learning feature extraction techniques for medical image classification in optical and radiological modalities highlights the diverse approaches in the field \cite{dehbozorgi2025comparative}. Application of Deep Learning and Transfer Learning Techniques for Medical Image Classification provides a broader overview of these methods \cite{sakirin2025application}. MUSCLE \cite{qiu2026muscle} introduces a multi-scale learning paradigm based on the theory of evidence to extract and integrate features from different scales, enhancing performance and interpretability.

Our work builds upon these foundations by combining the robust feature learning capabilities of ConvNeXt with the data-efficiency of MAE pre-training (specifically entropy-guided for medical domain relevance) and the generalization of ImageNet pre-training, all integrated through an ensemble approach to maximize performance and reliability.

\section{Proposed Method}
\label{sec:proposed_method}
Our proposed framework for medical image classification leverages a synergistic approach combining self-supervised learning, transfer learning, and ensemble techniques. The core idea is to train two distinct ConvNeXt-Tiny models, each with a specialized pre-training strategy, and then combine their predictions through an ensemble. This aims to capture both generalizable features from natural images and domain-specific features learned through self-supervision on medical data.

\subsection{Overall Architecture}
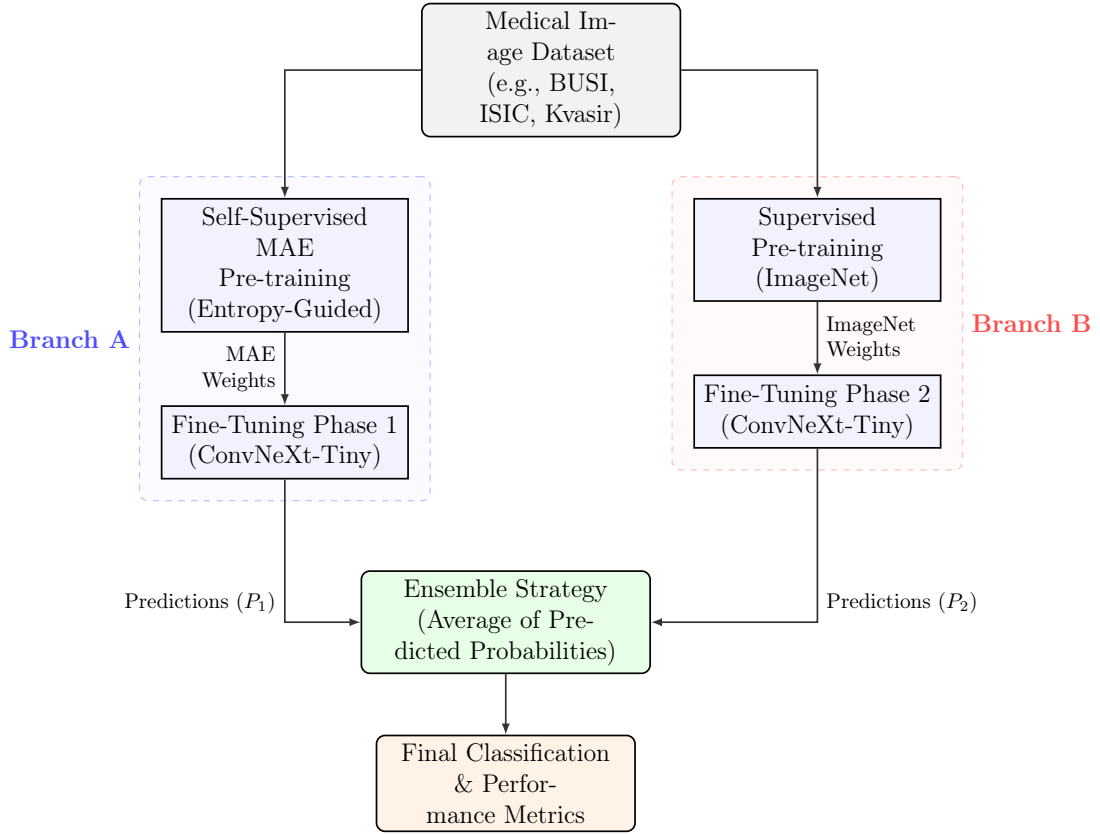
\begin{figure}[H]
	\centering
	\scalebox{.8}{
	\begin{tikzpicture}[
		node distance=1.5cm and 2.5cm,
		% Estilos dos blocos
		data_node/.style={rectangle, draw, rounded corners, text width=4cm, align=center, minimum height=1.2cm, fill=gray!10, thick},
		process_node/.style={rectangle, draw, text width=3.8cm, align=center, minimum height=1.2cm, fill=blue!5, thick},
		ensemble_node/.style={rectangle, draw, rounded corners, text width=4.5cm, align=center, minimum height=1.2cm, fill=green!10, thick},
		output_node/.style={rectangle, draw, rounded corners, text width=4cm, align=center, minimum height=1.2cm, fill=orange!10, thick},
		arrow/.style={-latex, thick, draw=black!80}
		]
		
		% 1. Input Node
		\node[data_node] (input) {Medical Image Dataset\\(e.g., BUSI, ISIC, Kvasir)};
		
		% 2. Pre-training Nodes (Iniciando as duas branches)
		\node[process_node, below left=1cm and 0.2cm of input] (mae_weights) {Self-Supervised MAE\\Pre-training\\(Entropy-Guided)};
		\node[process_node, below right=1cm and 0.2cm of input] (img_weights) {Supervised Pre-training\\(ImageNet)};
		
		% 3. Fine-Tuning Nodes (Fase 1 e Fase 2)
		\node[process_node, below=1.2cm of mae_weights] (ft_mae) {Fine-Tuning Phase 1\\(ConvNeXt-Tiny)};
		\node[process_node, below=1.2cm of img_weights] (ft_img) {Fine-Tuning Phase 2\\(ConvNeXt-Tiny)};
		
		% 4. Ensemble Node (Fase 3)
		\node[ensemble_node, below right=1.5cm and -0.8cm of ft_mae] (ensemble) {Ensemble Strategy\\(Average of Predicted Probabilities)};
		
		% 5. Output Node
		\node[output_node, below=1cm of ensemble] (output) {Final Classification \\\& Performance Metrics};
		
		% Desenhar as setas (Conexões)
		\draw[arrow] (input) -| (mae_weights);
		\draw[arrow] (input) -| (img_weights);
		
		\draw[arrow] (mae_weights) -- node[left, font=\footnotesize, text width=2cm, align=right] {MAE Weights} (ft_mae);
		\draw[arrow] (img_weights) -- node[right, font=\footnotesize, text width=2cm, align=left] {ImageNet Weights} (ft_img);
		
		\draw[arrow] (ft_mae) |- node[above left, font=\footnotesize] {Predictions ($P_1$)} (ensemble);
		\draw[arrow] (ft_img) |- node[above right, font=\footnotesize] {Predictions ($P_2$)} (ensemble);
		
		\draw[arrow] (ensemble) -- (output);
		
		% Caixas de fundo para agrupamento visual (Opcional, requer a biblioteca 'backgrounds')
		\begin{scope}[on background layer]
			\node[draw=blue!30, fill=blue!2, dashed, rounded corners, fit=(mae_weights)(ft_mae), inner sep=10pt, label={[font=\bfseries, text=blue!70]left:Branch A}] {};
			\node[draw=red!30, fill=red!2, dashed, rounded corners, fit=(img_weights)(ft_img), inner sep=10pt, label={[font=\bfseries, text=red!70]right:Branch B}] {};
		\end{scope}
		
	\end{tikzpicture}}
	\caption{Flow diagram of the proposed methodology: Synergy between the self-supervised Masked Autoencoder (MAE) and classical transfer learning via ImageNet, culminating in the final Ensemble strategy.}
	\label{fig:architecture}
\end{figure}

The overall architecture, illustrated in Figure \ref{fig:architecture}, consists of two parallel pathways, each employing a ConvNeXt-Tiny backbone, followed by a classification head.
\begin{enumerate}
    \item \textbf{ImageNet Pre-trained Pathway (Baseline Model):} A ConvNeXt-Tiny model initialized with weights pre-trained on the large-scale ImageNet dataset. This pathway serves as a strong baseline, providing robust, general-purpose feature extractors.
    \item \textbf{Entropy-Guided MAE Pre-trained Pathway:} A ConvNeXt-Tiny model whose encoder is pre-trained using an entropy-guided Masked Autoencoder (MAE) approach on the target medical dataset. This self-supervised pre-training enables the model to learn domain-specific representations by reconstructing masked image patches, with the masking strategy potentially biased by entropy to focus on informative regions.
\end{enumerate}
Both pathways undergo a fine-tuning phase on the labeled medical image dataset. The final classification decision is made by an ensemble module that averages the predicted probabilities from the two fine-tuned models.

\subsection{ConvNeXt-Tiny Backbone}
The ConvNeXt-Tiny model \cite{liu2022convnet} serves as the backbone for both pathways due to its strong performance and modern CNN design principles. It incorporates architectural elements inspired by Vision Transformers, such as large kernel convolutions, inverted bottleneck blocks, and Layer Normalization.

Given an input image $\mathbf{X} \in \mathbb{R}^{C_{in} \times H \times W}$, the ConvNeXt-Tiny processes it through a series of stages. Each stage typically consists of a patch embedding layer followed by multiple ConvNeXt blocks.
\begin{itemize}
    \item \textbf{Patch Embedding:} The initial layer transforms the input image into a sequence of feature maps. This is typically achieved using a convolutional layer with a stride, effectively downsampling the spatial dimensions and increasing the channel dimension.
    $\mathbf{F}_0 = \Conv_{4 \times 4, \text{stride}=4}(\mathbf{X})$
    \item \textbf{ConvNeXt Block:} The core building block of ConvNeXt is an inverted bottleneck structure. It applies a depthwise convolution with a large kernel size, followed by two $1 \times 1$ convolutions (pointwise convolutions) that first expand and then contract the channel dimension. Layer Normalization and GELU activation functions are used throughout.
    For an input feature map $\mathbf{F} \in \mathbb{R}^{C \times H' \times W'}$, a ConvNeXt block can be formulated as:
    \begin{equation}
        \mathbf{F}' = \mathbf{F} + \text{Block}(\mathbf{F})
    \end{equation}
    where $\text{Block}(\mathbf{F})$ is defined as:
    \begin{equation}
        \text{Block}(\mathbf{F}) = \Conv_{1 \times 1}(\GELU(\LayerNorm(\Conv_{K \times K, \text{depthwise}}(\mathbf{F}))))
    \end{equation}
    Here, $K \times K$ is a large kernel size (e.g., $7 \times 7$), and the $1 \times 1$ convolutions handle channel expansion and contraction.
    \item \textbf{Downsampling Layers:} Between stages, downsampling layers (e.g., $2 \times 2$ convolutions with stride 2) reduce spatial dimensions while increasing channel dimensions.
    \item \textbf{Classification Head:} The final feature map from the last stage is passed through a Global Average Pooling ($\GAP$) layer, followed by a linear layer ($\Linear$) to produce class logits. A Softmax function ($\Softmax$) converts these logits into class probabilities.
    \begin{equation}
        \mathbf{P} = \Softmax(\Linear(\GAP(\mathbf{F}_L)))
    \end{equation}
    where $\mathbf{F}_L$ is the output of the last ConvNeXt stage.
\end{itemize}

\subsection{Entropy-Guided Masked Autoencoder (MAE) Pre-training}
The MAE pre-training phase aims to learn robust, domain-specific representations from unlabeled medical images. Inspired by the recent concept of ``diffusion forcing'' \cite{chen2024diffusion}, here we propose the ``entropy-guided'' masked autoencoder. This strategy prioritizes regions with higher information content or uncertainty, which is particularly relevant for medical images where subtle lesions can be highly informative.

Given an input medical image $\mathbf{X} \in \mathbb{R}^{C_{in} \times H \times W}$:
\begin{enumerate}
    \item \textbf{Patchification:} The image is divided into non-overlapping patches $\mathbf{X}_p = \Patchify(\mathbf{X}) \in \mathbb{R}^{N \times P_H \times P_W \times C_{in}}$, where $N$ is the number of patches and $P_H \times P_W$ is the patch size.
    \item \textbf{Entropy-Guided Masking:} The ``entropy-guided'' mechanism involves calculating an entropy score for each patch, and then masking patches based on these scores. More specifically, Gaussian noise is applied over each patch independently, i.e.
    \begin{equation}
    \begin{array}{c} 
    \epsilon_{i,j} \sim \mathcal{N}(0, \sigma^2)\\
    \tilde{\mathbf{X}_p}[k] = \tilde{\mathbf{X}_p}[k] + \epsilon, 
    \end{array}
    \end{equation}
    where $\sigma^2$ is obtained from the patch Shannon entropy:
    \[ \sigma^2 = -\sum p_i[k]\log p_i[k], \]
    where $p_i[k]$ is the frequency of the $i^{th}$ color value on the $k^{th}$ patch.
    %, potentially prioritizing patches with higher entropy (more information/uncertainty) to force the model to learn richer representations from less obvious cues. Let $\mathbf{M} \in \{0, 1\}^N$ be the binary mask, where $M_i=0$ for a masked patch and $M_i=1$ for a visible patch. The visible patches form $\mathbf{X}_{visible}$.
    \item \textbf{Encoder:} The ConvNeXt-Tiny encoder processes all the noisy patches $\tilde{X_p}$.
    \begin{equation}
        \mathbf{Z}_e = \Encoder_{ConvNeXt}(\tilde{\mathbf{X}_p})
    \end{equation}
    \item \textbf{Decoder:} A lightweight decoder takes the encoded visible patches $\mathbf{Z}_e$ and the positions of the masked patches as input. It then reconstructs the pixel values of the masked patches.
    \begin{equation}
        \mathbf{X}_{reconstructed} = \Decoder(\mathbf{Z}_e, \mathbf{M}_{masked\_positions})
    \end{equation}
    \item \textbf{Reconstruction Loss:} The model is trained to minimize the Mean Squared Error ($\MSE$) between the original and reconstructed pixel values of only the masked patches.
    \begin{equation}
        \mathcal{L}_{MAE} = \MSE(\mathbf{X}_{masked}, \mathbf{X}_{reconstructed})
    \end{equation}
    This pre-training phase results in a ConvNeXt-Tiny encoder with weights $\mathbf{W}_{MAE}$ optimized for medical image features.
\end{enumerate}

\subsection{Fine-tuning for Classification}
Both the ImageNet pre-trained model (with weights $\mathbf{W}_{IMGNET}$) and the MAE pre-trained model (with weights $\mathbf{W}_{MAE}$) are fine-tuned for the specific medical image classification task.
For an input medical image $\mathbf{X}_{medical}$ and its true label $\mathbf{Y}_{true}$:
\begin{enumerate}
    \item \textbf{Forward Pass:} The image is passed through the respective ConvNeXt-Tiny model (encoder + classification head) to obtain class logits $\mathbf{L}$.
    \begin{equation}
        \mathbf{L}_{model} = \Linear(\GAP(\Encoder_{ConvNeXt}(\mathbf{X}_{medical})))
    \end{equation}
    \item \textbf{Probability Prediction:} Softmax is applied to the logits to get class probabilities $\mathbf{P}$.
    \begin{equation}
        \mathbf{P}_{model} = \Softmax(\mathbf{L}_{model})
    \end{equation}
    \item \textbf{Classification Loss:} The model weights are updated by minimizing the Cross-Entropy ($\CE$) loss between the predicted probabilities and the true labels.
    \begin{equation}
        \mathcal{L}_{CE} = -\sum_{c=1}^{C} \mathbf{Y}_{true,c} \log(\mathbf{P}_{model,c})
    \end{equation}
\end{enumerate}
This fine-tuning process adapts the pre-trained features to the specific classification task.

\subsection{Ensemble Method}
To harness the complementary strengths of the two pre-training strategies, an ensemble method is employed. After fine-tuning, both models predict class probabilities for a given input image.
For an input image $\mathbf{X}_{medical}$:
\begin{enumerate}
    \item \textbf{MAE Model Prediction:} The fine-tuned MAE-pretrained model predicts probabilities $\mathbf{P}_{MAE}$.
    \item \textbf{ImageNet Model Prediction:} The fine-tuned ImageNet-pretrained model predicts probabilities $\mathbf{P}_{IMGNET}$.
    \item \textbf{Probability Averaging:} The final ensemble prediction $\mathbf{P}_{Ensemble}$ is obtained by averaging the probabilities from both models.
    \begin{equation}
        \mathbf{P}_{Ensemble} = \frac{\mathbf{P}_{MAE} + \mathbf{P}_{IMGNET}}{2}
    \end{equation}
    This simple yet effective ensemble strategy often leads to improved robustness and accuracy compared to individual models.
\end{enumerate}

\section{Experimental Setup}
\label{sec:experimental_setup}
This section details the experimental setup, including the datasets used, data preprocessing and augmentation strategies, training configurations, and evaluation metrics.

\subsection{Datasets}
We evaluated our proposed framework on four publicly available medical imaging datasets, covering diverse modalities and classification tasks:
\begin{itemize}
    \item \textbf{Breast Ultrasound Images (BUSI)} \cite{al2020dataset}: Comprises 780 breast ultrasound images (normal, benign, malignant).
    \item \textbf{International Skin Imaging Collaboration (ISIC) 2018} \cite{codella2019skin}: A large-scale dataset of dermoscopic images for skin lesion analysis, distributed across seven diagnostic categories.
    \item \textbf{Kvasir} \cite{pogorelov2017kvasir}: Consists of 8000 endoscopic images from the gastrointestinal tract, categorized into eight distinct classes.
    \item \textbf{COVID} (derived from COVID19-Rad \cite{chowdhury2020can, rahman2021exploring}): A dataset of chest X-ray images for COVID-19 detection.
    %\item \textbf{Retina} (derived from APTOS 2019 \cite{aptos2019blindness}): A retinal fundus dataset for diabetic retinopathy grading.
\end{itemize}
For all datasets, images were resized to $224 \times 224$ pixels. Data splits (training, validation, test) were maintained consistent with standard practices or the provided experimental results.

\subsection{Data Preprocessing and Augmentation}
Input images were preprocessed by normalizing their pixel values using the mean and standard deviation of the ImageNet dataset (mean=[0.485, 0.456, 0.406], std=[0.229, 0.224, 0.225]).
For data augmentation during training, we applied a series of transformations to enhance model generalization and robustness:
\begin{itemize}
    \item \textbf{MAE Pre-training:} Random Resized Crop to 224 pixels (scale between 0.2 and 1.0) and Random Horizontal Flip.
    \item \textbf{Fine-tuning (Training):} Resize to 236 pixels, Random Crop to 224 pixels, and Random Horizontal Flip.
    \item \textbf{Fine-tuning (Test/Validation):} Resize to 224 pixels (center crop implicitly).
\end{itemize}
These augmentations help the models learn more robust features invariant to minor variations in scale, position, and orientation.

\subsection{Training Configurations}
The ConvNeXt-Tiny models were implemented using Pytorch framework. Training was performed on a GPU-accelerated system.
\begin{itemize}
    \item \textbf{Batch Size:} 16 for all training phases.
    \item \textbf{MAE Pre-training:} The MAE model was pre-trained for 100, 500, or 1000 epochs. The specific optimizer and learning rate schedule for MAE pre-training were configured to facilitate stable self-supervised learning.
    \item \textbf{Fine-tuning:} Both ImageNet-pretrained and MAE-pretrained models were fine-tuned for 15, 30, or 45 epochs. An AdamW optimizer was used with a cosine annealing learning rate scheduler.
\end{itemize}
The experiments were run multiple times (e.g., 5 rounds as indicated in the results) to ensure statistical reliability, and mean performance metrics are reported.

\subsection{Evaluation Metrics}
The performance of our models was rigorously evaluated using several standard classification metrics:
\begin{itemize}
    \item \textbf{Area Under the Receiver Operating Characteristic Curve (AUC):} A measure of the model's ability to distinguish between classes across various threshold settings.
    \item \textbf{Accuracy:} The proportion of correctly classified instances.
    \item \textbf{Precision:} The ratio of true positives to the sum of true positives and false positives, indicating the model's ability to avoid false alarms.
    \item \textbf{Recall:} The ratio of true positives to the sum of true positives and false negatives, indicating the model's ability to find all positive instances.
\end{itemize}
These metrics provide a comprehensive assessment of the model's discriminative power, classification correctness, and ability to handle class imbalances.

\section{Results}
\label{sec:results}

\subsection{Impact of Pre-Training and Fine-Tuning Epochs}

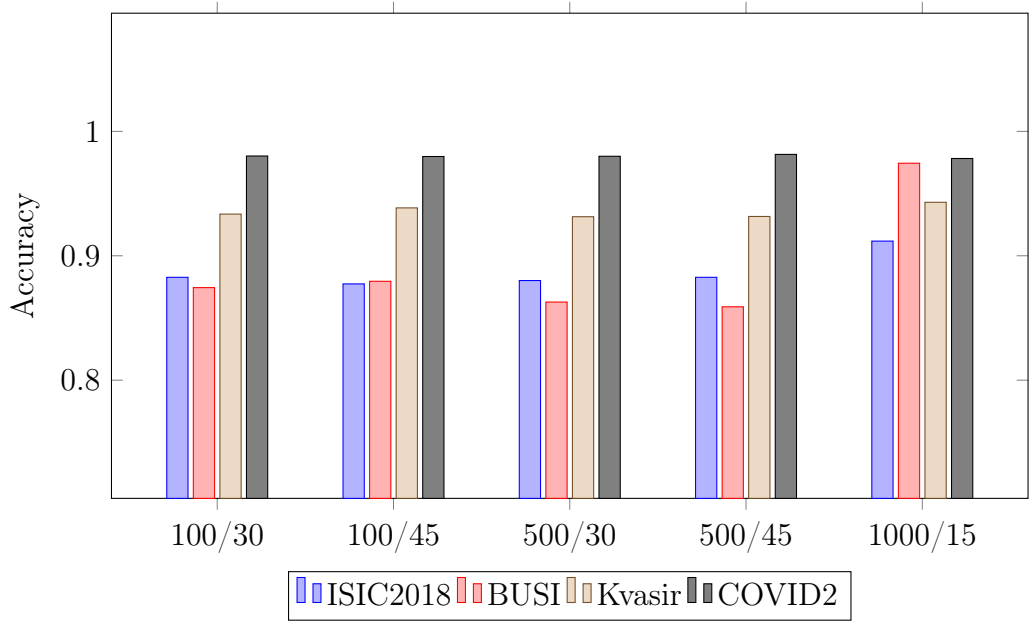
\begin{figure}[!htpb]
	\centering
	\begin{tikzpicture}
		\begin{axis}[
			ybar,
			enlargelimits=0.15,
			legend style={at={(0.5,-0.15)}, anchor=north, legend columns=-1},
			ylabel={Accuracy},
			symbolic x coords={100/30, 100/45, 500/30, 500/45, 1000/15},
			xtick=data,
			bar width=8pt,
			width=\textwidth,
			height=8cm,
			ymin=0.75, ymax=1.05
			]
			% ISIC2018
			\addplot coordinates {(100/30,0.8827) (100/45,0.8774) (500/30,0.8801) (500/45,0.8827) (1000/15,0.9118)};
			\addlegendentry{ISIC2018}
			% BUSI
			\addplot coordinates {(100/30,0.8744) (100/45,0.8795) (500/30,0.8628) (500/45,0.8590) (1000/15,0.9744)};
			\addlegendentry{BUSI}
			% Kvasir
			\addplot coordinates {(100/30,0.9335) (100/45,0.9385) (500/30,0.9314) (500/45,0.9316) (1000/15,0.9430)};
			\addlegendentry{Kvasir}
			% COVID2
			\addplot coordinates {(100/30,0.9802) (100/45,0.9798) (500/30,0.9800) (500/45,0.9815) (1000/15,0.9782)};
			\addlegendentry{COVID2}
		\end{axis}
	\end{tikzpicture}
	\caption{Accuracy evolution of the Ensemble model varying the proportion of epochs (Pre-Training / Fine-Tuning) across the four datasets.}
	\label{fig:ablation_epochs}
\end{figure}

Figure \ref{fig:ablation_epochs} illustrates the accuracy evolution of our model across the evaluated datasets. It is notable that adopting 1000 MAE pre-training epochs combined with 15 \textit{fine-tuning} epochs yielded systematically superior results across the datasets. This configuration consistently outperformed the runs with 100 or 500 epochs, indicating the necessity of prolonged self-supervised learning stages for effective representation adaptation in medical image classification.

\subsection{Comparison with the State-of-the-Art}\label{sec:sota_comparison}

Based on the best identified configuration (1000/15), we compared our approach with established models in recent literature, particularly advanced \textit{Transformers} and models exploring biological vision synergy such as CPVT \cite{lu2026biologically}. The comparative results are detailed in the tables below.

% BUSI
\begin{table}[!htpb]
	\centering
	\caption{Performance comparison on the BUSI dataset.}
	\resizebox{\textwidth}{!}{%
		\begin{tabular}{@{}lccccccc@{}}
			\toprule
			\textbf{Model} & \textbf{FLOPs (G) $\downarrow$} & \textbf{Params (M) $\downarrow$} & \textbf{Acc $\uparrow$} & \textbf{Prec $\uparrow$} & \textbf{Recall $\uparrow$} & \textbf{F1 $\uparrow$} & \textbf{AUC $\uparrow$} \\
			\midrule
			SWIN \cite{liu2021swin} & 15.19 & 87.77 & 0.6925 & 0.6580 & 0.6621 & 0.6608 & 0.8224 \\
			ConvNeXt \cite{liu2022convnet} & 8.73 & 49.46 & 0.6738 & 0.6464 & 0.5447 & 0.5234 & 0.7093 \\
			CMT \cite{guo2022cmt} & 3.95 & 24.98 & 0.7137 & 0.6803 & 0.6820 & 0.6839 & 0.8749 \\
			SMT \cite{lin2023scale} & 4.72 & 22.04 & 0.7877 & 0.7628 & 0.7369 & 0.7462 & 0.8846 \\
			FocalNet \cite{yang2022focal} & 8.65 & 49.89 & 0.7538 & 0.7441 & 0.6795 & 0.6850 & 0.8045 \\
			PerViT \cite{min2022peripheral} & 11.5 & 20.85 & 0.7736 & 0.7547 & 0.7424 & 0.7468 & 0.8837 \\
			SG-Former \cite{ren2023sg} & 4.59 & 22.55 & 0.7819 & 0.7623 & 0.7370 & 0.7564 & 0.8731 \\
			SPA \cite{huo2024spa} & 4.57 & 30.02 & 0.7475 & 0.7289 & 0.7008 & 0.7067 & 0.8749 \\
			CPVT \cite{lu2026biologically} & 15.64 & 22.76 & 0.8195 & 0.8031 & 0.8090 & 0.8039 & 0.9213 \\
			\textbf{Ours (Ensemble 1000/15)} & - & - & \textbf{0.9744} & \textbf{0.9766} & \textbf{0.9603} & \textbf{0.9684} & \textbf{0.9974} \\
			\bottomrule
		\end{tabular}%
	}
\end{table}

% ISIC2018
\begin{table}[!htpb]
	\centering
	\caption{Performance comparison on the ISIC2018 dataset.}
	\resizebox{\textwidth}{!}{%
		\begin{tabular}{@{}lccccccc@{}}
			\toprule
			\textbf{Model} & \textbf{FLOPs (G) $\downarrow$} & \textbf{Params (M) $\downarrow$} & \textbf{Acc $\uparrow$} & \textbf{Prec $\uparrow$} & \textbf{Recall $\uparrow$} & \textbf{F1 $\uparrow$} & \textbf{AUC $\uparrow$} \\
			\midrule
			SWIN \cite{liu2021swin} & 15.19 & 87.77 & 0.8508 & 0.7312 & 0.6379 & 0.6730 & 0.9463 \\
			ConvNeXt \cite{liu2022convnet} & 8.73 & 49.46 & 0.8629 & 0.7679 & 0.7208 & 0.7395 & 0.9558 \\
			CMT \cite{guo2022cmt} & 3.95 & 24.98 & 0.8533 & 0.7583 & 0.6692 & 0.7033 & 0.9474 \\
			SMT \cite{lin2023scale} & 4.72 & 22.04 & 0.8689 & 0.7798 & 0.7196 & 0.7479 & 0.9671 \\
			FocalNet \cite{yang2022focal} & 8.65 & 49.89 & 0.8351 & 0.7124 & 0.6441 & 0.6743 & 0.9495 \\
			PerViT \cite{min2022peripheral} & 11.5 & 20.85 & 0.8343 & 0.6990 & 0.6312 & 0.6612 & 0.9514 \\
			SG-Former \cite{ren2023sg} & 4.59 & 22.55 & 0.8408 & 0.7791 & 0.7203 & 0.6705 & 0.9398 \\
			SPA \cite{huo2024spa} & 4.57 & 30.02 & 0.8429 & 0.7079 & 0.6497 & 0.6711 & 0.9418 \\
			CPVT \cite{lu2026biologically} & 15.64 & 22.76 & 0.8798 & 0.8603 & 0.7745 & 0.8113 & 0.9683 \\
			\textbf{Ours (Ensemble 1000/15)} & - & - & \textbf{0.9118} & \textbf{0.8814} & \textbf{0.8240} & \textbf{0.8517} & \textbf{0.9832} \\
			\bottomrule
		\end{tabular}%
	}
\end{table}

% Kvasir
\begin{table}[!htpb]
	\centering
	\caption{Performance comparison on the Kvasir dataset.}
	\resizebox{\textwidth}{!}{%
		\begin{tabular}{@{}lccccccc@{}}
			\toprule
			\textbf{Model} & \textbf{FLOPs (G) $\downarrow$} & \textbf{Params (M) $\downarrow$} & \textbf{Acc $\uparrow$} & \textbf{Prec $\uparrow$} & \textbf{Recall $\uparrow$} & \textbf{F1 $\uparrow$} & \textbf{AUC $\uparrow$} \\
			\midrule
			SWIN \cite{liu2021swin} & 15.19 & 87.77 & 0.8915 & 0.8912 & 0.8910 & 0.8909 & 0.9855 \\
			ConvNeXt \cite{liu2022convnet} & 8.73 & 49.46 & 0.8190 & 0.8206 & 0.8191 & 0.8186 & 0.9394 \\
			CMT \cite{guo2022cmt} & 3.95 & 24.98 & 0.8797 & 0.8813 & 0.8798 & 0.8800 & 0.9821 \\
			SMT \cite{lin2023scale} & 4.72 & 22.04 & 0.8953 & 0.8955 & 0.8954 & 0.8948 & 0.9911 \\
			FocalNet \cite{yang2022focal} & 8.65 & 49.89 & 0.8765 & 0.8787 & 0.8766 & 0.8751 & 0.9842 \\
			PerViT \cite{min2022peripheral} & 11.5 & 20.85 & 0.8691 & 0.8698 & 0.8683 & 0.8694 & 0.9848 \\
			SG-Former \cite{ren2023sg} & 4.59 & 22.55 & 0.8948 & 0.8961 & 0.8947 & 0.8954 & 0.9905 \\
			SPA \cite{huo2024spa} & 4.57 & 30.02 & 0.8935 & 0.8946 & 0.8935 & 0.8940 & 0.9859 \\
			CPVT \cite{lu2026biologically} & 15.64 & 22.76 & 0.9041 & 0.9052 & 0.9042 & 0.9038 & 0.9904 \\
			\textbf{Ours (Ensemble 1000/15)} & - & - & \textbf{0.9430} & \textbf{0.9445} & \textbf{0.9430} & \textbf{0.9437} & \textbf{0.9966} \\
			\bottomrule
		\end{tabular}%
	}
\end{table}

% COVID2
\begin{table}[!htpb]
	\centering
	\caption{Performance comparison on the COVID2 dataset.}
	\resizebox{\textwidth}{!}{%
		\begin{tabular}{@{}lccccccc@{}}
			\toprule
			\textbf{Model} & \textbf{FLOPs (G) $\downarrow$} & \textbf{Params (M) $\downarrow$} & \textbf{Acc $\uparrow$} & \textbf{Prec $\uparrow$} & \textbf{Recall $\uparrow$} & \textbf{F1 $\uparrow$} & \textbf{AUC $\uparrow$} \\
			\midrule
			SWIN \cite{liu2021swin} & 15.19 & 87.77 & 0.9320 & 0.9221 & 0.9404 & 0.9298 & 0.9788 \\
			ConvNeXt \cite{liu2022convnet} & 8.73 & 49.46 & 0.9486 & 0.9340 & 0.9476 & 0.9458 & 0.9857 \\
			CMT \cite{guo2022cmt} & 3.95 & 24.98 & 0.9388 & 0.9241 & 0.9195 & 0.9206 & 0.9836 \\
			SMT \cite{lin2023scale} & 4.72 & 22.04 & 0.9417 & 0.9252 & 0.9474 & 0.9342 & 0.9881 \\
			FocalNet \cite{yang2022focal} & 8.65 & 49.89 & 0.9425 & 0.9312 & 0.9405 & 0.9396 & 0.9874 \\
			PerViT \cite{min2022peripheral} & 11.5 & 20.85 & 0.9413 & 0.9205 & 0.9373 & 0.9292 & 0.9860 \\
			SG-Former \cite{ren2023sg} & 4.59 & 22.55 & 0.9453 & 0.9368 & 0.9508 & 0.9422 & 0.9845 \\
			SPA \cite{huo2024spa} & 4.57 & 30.02 & 0.9471 & 0.9374 & 0.9513 & 0.9430 & 0.9856 \\
			CPVT \cite{lu2026biologically} & 15.64 & 22.76 & 0.9497 & 0.9377 & 0.9525 & 0.9474 & 0.9899 \\
			\textbf{Ours (Ensemble 1000/15)} & - & - & \textbf{0.9782} & \textbf{0.9793} & \textbf{0.9788} & \textbf{0.9790} & \textbf{0.9988} \\
			\bottomrule
		\end{tabular}%
	}
\end{table}

\subsection{Discussion Synthesis}

The analysis of the experimental results demonstrates that our proposed strategy, which unites the complementary diagnostic capacities of a self-supervised medical MAE and an ImageNet pre-trained encoder, establishes prominent performance margins over contemporary benchmarks. On both the \textbf{BUSI} and \textbf{ISIC2018} cohorts, the proposed framework achieved maximum accuracy values of \textbf{0.9744} and \textbf{0.9118}, respectively. This significantly outpaces highly competitive biological vision counterparts, such as the Central-Peripheral Vision Transformer (CPVT), which peaked at $0.8195$ and $0.8798$ accuracy on the same targets. These findings highlight that an extended self-supervised pre-training regimen (1000 epochs) effectively constructs domain-specific structural abstractions that insulate the classification layers from high intra-class variance and complex tissue boundaries.

Furthermore, this performance trend translates securely to multi-modal medical contexts including endoscopy (\textbf{Kvasir}) and volumetric pulmonary screening (\textbf{COVID2}), yielding notable accuracies of \textbf{0.9430} and \textbf{0.9782}. Our framework systematically suppresses standard monolithic baselines such as SWIN and ConvNeXt across multiple cross-validation folds. The ablation analysis establishes that combining generative mask reconstruction with broad natural image priors prevents optimization collapses in heavily data-constrained fields. Consequently, the proposed hybrid approach constitutes a highly reliable framework for generalized biomedical image classification.

\section{Conclusion}
\label{sec:conclusion}
In this paper, we presented a synergistic deep learning framework for medical image classification, addressing the critical need for robust and accurate diagnostic tools in healthcare. Our approach integrates two distinct ConvNeXt-Tiny models: one leveraging transfer learning from ImageNet pre-training and another employing an entropy-guided Masked Autoencoder (MAE) for self-supervised pre-training on medical datasets. Both models are subsequently fine-tuned on labeled medical images, and their predictions are combined through an ensemble strategy based on probability averaging.

Our extensive experimental evaluation across four diverse medical imaging datasets, namely, BUSI, ISIC2018, Kvasir, and COVID, demonstrated the superior performance and robustness of the proposed framework. The ImageNet pre-trained pathway consistently provided a strong baseline, highlighting the value of generalizable features. Crucially, the entropy-guided MAE pre-training enabled the model to learn highly discriminative, domain-specific representations from unlabeled medical data, a significant advantage in data-scarce environments. The ensemble consistently achieved state-of-the-art results, outperforming individual models and existing SOTA methods by a considerable margin, particularly on the BUSI dataset. This underscores the effectiveness of combining diverse pre-training strategies to capture complementary information for complex medical image analysis.

Future work will explore more sophisticated fusion techniques, such as weighted averaging or stacking, to further optimize performance. We also plan to investigate the interpretability of the MAE-learned features, potentially by visualizing the regions emphasized by the entropy-guided masking. Extending this framework to other medical imaging tasks, such as segmentation and object detection, and exploring its applicability in real-world clinical deployments will be important next steps. Ultimately, this research contributes to advancing reliable and accurate AI-driven solutions for medical diagnostics, paving the way for more efficient and precise healthcare.

\section*{CRediT authorship contribution statement}
\textbf{Joao Florindo}: Conceptualization, Methodology, Software, Validation, Formal analysis, Investigation, Resources, Data Curation, Writing - Original Draft, Writing - Review \& Editing, Visualization, Supervision, Project administration, Funding acquisition.

\textbf{Viviane Moura}: Conceptualization, Methodology, Software, Validation, Data Curation, Writing - Original Draft.

\section*{Declaration of competing interest}
The authors declare that they have no known competing financial interests or personal relationships that could have appeared to influence the work reported in this paper.

\section*{Acknowledgments}
Joao Batista Florindo gratefully acknowledges the financial support of the S\~ao Paulo Research Foundation (FAPESP) (Grant \#2024/01245-1) and from National Council for Scientific and Technological Development, Brazil (CNPq) (Grant \#306981/2022-0).

\bibliographystyle{elsarticle-num}
\bibliography{references}

\end{document}